\documentclass[a4paper,12pt]{article}
\usepackage{amsmath}
\usepackage{amssymb}
\usepackage{latexsym}
\usepackage{lscape}
\usepackage{pdflscape}
\usepackage{hyperref}
\usepackage[dvipsnames]{xcolor}
\usepackage{amsfonts}
\usepackage{amsmath}
\usepackage{latexsym}
\usepackage{enumitem}
\usepackage{graphics}
\usepackage{bbm}
\usepackage{tensor}
\usepackage[framemethod=tikz,skipabove=1pt,innertopmargin = 0pt,skipbelow = 1pt,innerbottommargin = 0pt,innerleftmargin = 0pt,innerrightmargin = 0pt,leftmargin = 0pt,rightmargin = 0pt]{mdframed} 
\usepackage{hyperref}
\newcommand{\be}{\begin{eqnarray}}
\newcommand{\ea}{\end{eqnarray}}
\newcommand{\ben}{\begin{equation*}}
\newcommand{\een}{\end{equation*}}
\newcommand{\bean}{\begin{eqnarray*}}
\newcommand{\eean}{\end{eqnarray*}}
\def\bal#1\eal{\begin{align}#1\end{align}}
\newcommand{\ph}{\phantom}
\newcommand{\bsub}{\begin{subequations}}
\newcommand{\esub}{\end{subequations}}

\newcommand{\disfrac}[1][2]{\displaystyle\frac}

\newcommand{\non}{\nonumber}

\newcommand{\pound}{\emph{\textsterling}}

\newcommand{\drm}{\textrm{d}}
\newcommand*\xbar[1]{%
  \hbox{%
    \vbox{%
      \hrule height 0.5pt 
      \kern0.5ex
      \hbox{%
        \kern-0.1em
        \ensuremath{#1}%
        \kern-0.1em
      }%
    }%
  }%
}
\newcommand{\bmdf}{\begin{mdframed}[hidealllines=true,backgroundcolor=gray!20]}  
\newcommand{\emdf}{\end{mdframed}}

\numberwithin{equation}{section}
\begin{document}

\title{Noether analysis of Scalar--Tensor Cosmology}
\vspace{1cm}
\author{\textbf{Petros A. Terzis}\thanks{pterzis@phys.uoa.gr},\, \textbf{N. Dimakis}\thanks{nsdimakis@gmail.com}\,, \textbf{T. Christodoulakis}\thanks{tchris@phys.uoa.gr}\\
{\it Nuclear and Particle Physics Section, Physics
Department,}\\{\it University of Athens, GR 157--71 Athens}}

\date{}
\maketitle
\begin{abstract}
A scalar--tensor theory of gravity, containing an arbitrary coupling function $F(\phi)$ and a general potential $V(\phi)$, is considered in the context of a spatially flat FLRW model. The use of reparametrization invariance enables a particular lapse parametrization in which the mini--superspace metric completely specifies the dynamics of the system. A requirement of existence of the maximal possible number of autonomous integrals of motion is imposed. This leads to a flat mini--superspace metric realized by a particular relation between the coupling function and the potential. The space of solutions is completely described in terms of the three autonomous integrals of motion constructed by the Killing fields of the mini--supermetric and an additional rheonomous emanating from the homothetic field. The solutions contain the arbitrary function which remains after the imposition of the relation between $F(\phi)$ and $V(\phi)$. To exemplify the use of the general results, we select some particular cases and study their physical implications through an effective energy--momentum tensor, which tends out to be that of a perfect fluid.
\end{abstract}

\section{Introduction}
In the last fifteen years a groundbreaking discovery has altered the way we view our universe; namely observations show that the universe is not only expanding but it is also accelerating \cite{Perlmutter:1998,Riess:1998,Riess:1998dv}. This fact gave birth to a plethora of propositions for explaining it; only to list a few: quintessence models \cite{Ratra:1987rm,Tsujikawa:2013fta}, which invoke an evolving canonical scalar field with a potential; Chameleon fields in which the scalar field couples to the baryon energy density and is homogeneous \cite{Khoury:2003rn,Lombriser:2014dua}; a scalar field with
a non-canonical kinetic term, known as K-essence \cite{Chiba:1999ka,ArmendarizPicon:2000ah} based on earlier work of
K-inflation \cite{ArmendarizPicon:1999rj}; Chaplygin gases, which attempt to unify dark energy and dark matter under one roof by allowing for a fluid with an equation of state which evolves between the two \cite{Kamenshchik:2001cp,Bilic:2001cg,Bento:2002ps}; phantom dark energy \cite{Caldwell:2003vq} or even direct anthropic arguments \cite{Linde:1984ir,Weinberg:1988cp,Efstathiou:1990xe}; for a comprehensive review see \cite{Sahni:2004ai,Copeland:2006wr}.

Another big field of research is devoted to modified theories of gravity and specifically to the scalar--tensor case; which is the subject of the current work. Scalar--tensor theories, with a non minimal coupling, are considered as the most general, since they incorporate a major amount of other theories. It is well known that $f(R)$ gravity theories are equivalent to many scalar--tensor cases, with the derivative of the function $f(R)$ playing the role of the Brans--Dicke scalar \cite{Higgs1959,Chiba2003,Teyssandier:1983zz,Wands1994}; fourth-order gravity theory \cite{Higgs1959,Whitt1984} are also equivalent to a scalar tensor theory and there is even a big analogy among the $f (R)$--gravity with torsion and scalar--tensor theories with torsion, as discussed, for example, in \cite{German1985,Sung-Won1986} (for a review of all of them see \cite{Capozziello:2011et}).

The use of Noether symmetries in minisuperspace, either in classical or in quantum level, is not new. This approach for classical Bianchi cosmologies has been to the best of our knowledge, initiated in \cite{Capozziello:1996ay} and then used in \cite{Capozziello:1999xr,Cotsakis:1998zk}; while work on the subject has been revived from numerous authors \cite{tsamp1,Basilakos:2011rx,Vakili:2011uz,PintoNeto:2012ug,Capozziello:2012hm,Sarkar:2012sx}.

The common feature of all the above works is that they were dealing with systems described by singular Lagrangians, since all of them admit a time reparametrization invariance. In \cite{CDT2014} the symmetry treatment of such Lagrangians was addressed and it was shown how one can find all the Noether symmetries possessed by these systems. The result is that we have to extend the infinitesimal criterion of symmetry in such a way that it includes the constraint that arises from the reparametrization invariance. This method was used in \cite{Christodoulakis:2001um,tchris_sch,Christodoulakis:2013sya,Christodoulakis:2014wba} for the quantization of various minisuperspace models and in \cite{Dimakis:2013oza} where a Noether analysis of FRLW cosmology in the context of $f(R)$--gravity was performed, resulting in the discovery of several exact new solutions.

In the present work we use the method developed in \cite{CDT2014}, to
investigate a general non--minimal coupling for a scalar field  $\phi$ with gravity, which is proportional to the Ricci scalar $R$, see \eqref{action} below, embedded in a Friedmann--Lema\^itre--Robertson--Walker (FRLW) spacetime. The strategy we follow is to demand a maximal number of Noether symmetries of the action \eqref{action} in order to find the general solution for the scale factor $a(t)$, the coupling function $F(\phi)$, the potential $V(\phi)$ and the scalar field $\phi(t)$.

In order to infer the physical properties of the solutions we obtain, we start from the known duality between scalar fields and perfect fluids \cite{Madsen1, Madsen2, Pimentel}. The usual line of thought is to try to interpret the energy momentum tensor of the scalar field as an energy momentum tensor of a perfect fluid \cite{Unnikrishnan:2010ag,Christopherson:2008ry,DiezTejedor:2005fz, Arroja:2010wy}; of course this duality must be taken with caution e.g. at the level of the Lagrangian formulation problems may arise as recently noted \cite{Faraoni:2012hn}.

We, on the other hand, choose to make a slightly different identification; we rewrite the field equations of the scalar--tensor theory, as in General Relativity, i.e. $G_{ij}=T_{ij}$ and interpret the right hand side as the energy momentum tensor of an imperfect fluid. The nice outcome is that in the general case the imperfect fluid is actually a perfect one. In order to pick up physically acceptable perfect fluid solutions, one must demand a sort of energy conditions.

The structure of the paper is the following. In Sec. 2, we set up the field equations, perform the Noether analysis and calculate the general solutions. In Sec. 3, we calculate the parameters that characterize the universe expansion (Hubble, deceleration and jerk) and establish the correspondence between a scalar field and a perfect fluid. In Sec. 4, we present a number of special solutions, among them one that obeys the major energy conditions and describes an expanding universe suffering from a cosmic jerk (a deceleration epoch followed by an accelerating one). Finally Sec. 5 is devoted to discussion.

\section{Noether analysis and general solutions}
\subsection{Background geometry and minisuperspace}

Let us consider a FLRW space--time, which describes a homogeneous and spatially flat universe, i.e.
\bal\label{FLRW_metric}
\drm s^2=-N(u)^2 \drm u^2+a(u)^2\left( \drm r^2+r^2\, \drm \Omega^2 \right) ,
\eal
where $ \drm \Omega^2=\drm\theta^2+\sin^2\theta\, \drm\varphi^2$ and $N(u)$ is the lapse function which will play an essential role in the development of our treatment of the problem.

The action that describes the non--minimal coupling between gravity and the scalar field $\phi$ is taken as
\bal\label{action}
S=\int\!\sqrt{-g}\left( F(\phi)R+\frac{\epsilon}{2}g_{mn}\phi^{;m}\phi^{;n}-V(\phi) \right) \drm^4 x,
\eal
where $F(\phi)$ a function that represents the coupling, the constant $\epsilon$ equals $\epsilon=\pm 1$ allowing ghost fields ($\epsilon=+1$), $V(\phi)$ is a self--interaction potential and $R$ is the Ricci scalar.

In order to find the field equations along with the equation that the scalar field $\phi$ obeys, we must vary the action \eqref{action} with respect to $g_{ij}$ and $\phi$ respectively. The result is
\bsub\label{EoM}
\bal
F(\phi)\left(R_{ij}-\frac{1}{2}Rg_{ij} \right)-\nabla_i\nabla_j F(\phi)+g_{ij}\Box F(\phi)&=T_{ij}, \label{EoM1}\\
 \Box\phi+\epsilon V'(\phi)&=\epsilon F'(\phi)R \label{EoM2},
\eal
\esub
where
\bal\label{Tij}
T_{ij}=-\frac{\epsilon}{2}\nabla_i\phi \nabla_j\phi+\frac{1}{4}g_{ij}\left(\epsilon \nabla^k\phi\nabla_k\phi-2V(\phi) \right),
\eal
is the energy--monentum tensor, $\Box=\nabla_k\nabla^k$ is the Laplace--Beltrami operator,  $\nabla_k$ is the covariant derivative and the prime $'$ on a function, denotes the derivative with respect to its argument.

The Lagrangian treatment of the problem begins by inserting the values of $g_{ij}$ from \eqref{FLRW_metric} in \eqref{action}. The resulting Lagrangian is given by
\bal\label{Lag}
L=\frac{1}{2N}\,G_{\alpha\beta} (x^\alpha) x'^\alpha\, x'^\beta-N\,U(x^\alpha), \quad  G_{\alpha\beta} =
\begin{pmatrix}
-12a F & -6 a^2 F_\phi\\
-6 a^2 F_\phi & -\epsilon a^3
\end{pmatrix},
\quad U=a^3 V,
\eal
where $x^\alpha=(a,\phi)$\footnote{Greek indices take the values $1,2$ while the range of the English ones is $1\dots 4$}, the subscript $\phi$ indicates the derivative with respect to $\phi$ and $G_{\alpha\beta}$ is the mini--supermetric of our problem. It is an essential requirement in differential geometry to check, that the field equations \eqref{EoM} and the Euler--Lagrange equations yielding from the reduced Lagrangian \eqref{Lag} are equivalent; something that is true  in our case.

\subsection{Noether Symmetries}
One way to solve the equations of motion \eqref{EoM} resulting from \eqref{Lag} is to search for the Noether symmetries that the system possesses. The significant feature of this Lagrangian is that it is singular, since there is no $N'$ term; thus, in order to find its Noether symmetries we must take this fact into consideration.

The correct way of treating this sort of Lagrangians, in order to acquire all their Noether symmetries, was exhibited in \cite{CDT2014}; the result is that the Noether symmetries correspond to the conformal Killing fields of both $G_{\alpha\beta}$ and $U(x^\alpha)$ with opposite conformal factors, i.e.
\bal
\pound_\xi G_{\alpha\beta}=\omega(x^\alpha)\,G_{\alpha\beta}, \quad \pound_\xi U =-\omega (x^\alpha) \,U(x^\alpha)
\eal

The freedom of time re--parametrization, allows us to redefine the lapse function $N(t)$ in such a way so that the potential $U(x^\alpha)$ becomes constant; the recipe is to define a new lapse $\xbar{N}= N/U(x^\alpha)$, which in turn scales the mini--supermetric to $\xbar{G}_{\alpha\beta}= U(x^\alpha)G_{\alpha\beta}$. In this parametrization the symmetries of \eqref{Lag} corresponding to integrals of motion are constructed by all the Killing fields of the scaled supermetric $\xbar{G}_{\alpha\beta}= U(x^\alpha)G_{\alpha\beta}$; Additionally, its homothetic field (which is a Lie-point symmetry of the equations of motion) can be used to define a rheonomous integral of motion, the details are explained in \cite{CDT2014}.

The scaled mini--supermetric $G_{\alpha\beta}$ reads (we drop the bars hereafter)
\bal\label{scGdd}
G_{\alpha\beta} =a^3\,V
\begin{pmatrix}
-12a F & -6 a^2 F_\phi\\
-6 a^2 F_\phi & -\epsilon a^3
\end{pmatrix},
\eal
while the corresponding Ricci scalar is proportional to
\bal\label{Ricci}
-2 F V_\phi^2\left(\epsilon  F-3 F_\phi^2\right)+\non\\
V \left(-6 F_\phi^3 V_\phi+F F_\phi \left(V_\phi \left(6 F_{\phi\phi}+\epsilon \right)-6 F_\phi V_{\phi\phi}\right)+2 \epsilon  F^2 V_{\phi\phi}\right)+\non\\
2 \epsilon  V^2 \left(F_\phi^2-2 F F_{\phi\phi}\right).
\eal

\subsubsection{Flat minisuperspace}

The proportionality factor of \eqref{Ricci} is a particular function of $a$. Thus, if one wants to have the maximum number of Noether symmetries, the only viable case is for the Ricci scalar to be zero, since it can not be a non zero constant. Therefore, one is led to the nihilism of the above expression, which can be achieved if $F(\phi),V(\phi)$ are assumed to satisfy
\bal\label{FVpar}
F(\phi)=\frac{1}{4}h^2(\phi),\, V(\phi)=e^{f(\phi)}h^4(\phi), \quad \text{where} \quad 3h'^2-\lambda^2h^2 f'^2=\epsilon.
\eal

The functions $f(\phi),h(\phi)$ are arbitrary and $\lambda$ is a constant.

In order to calculate the form of the Killing fields $\xi^\alpha$ and the homothetic field $\eta^\alpha$ of the scaled mini--supermetric we bring it to a diagonal form
\bal
G_{\alpha\beta}=64\exp\left( 2\sqrt{3}w+f(\phi) \right)
\begin{pmatrix}
-1 & 0\\
0 & \lambda\,f'^2(\phi)
\end{pmatrix}
\eal
 with the aid of the transformation
\bal\label{trans}
a=\frac{1}{\sqrt{F(\phi)}}e^{w/\sqrt{3}}.
\eal

The resulting fields are
\bsub
\bal
\xi_{(1)}&=-\frac{1}{2}\exp \frac{ \left(2\sqrt{3\lambda}-1\right) \left( -w+\sqrt{\lambda}f(\phi) \right) }{2\sqrt{\lambda}}\left( \partial_w-\frac{1}{\sqrt{\lambda} f'(\phi)}\partial_\phi \right)\\
\xi_{(2)}&=\frac{1}{2}\exp \frac{ -\left(2\sqrt{3\lambda}+1\right) \left( w+\sqrt{\lambda}f(\phi) \right) }{2\sqrt{\lambda}}\left( \partial_w+\frac{1}{\sqrt{\lambda} f'(\phi)}\partial_\phi \right)\\
\xi_{(3)}&=-\frac{1}{2}\partial_w+\frac{\sqrt{3}}{f'(\phi)}\partial_\phi\\
\eta&=\frac{1}{2\sqrt{3}}\partial_w.
\eal
\esub

From the above fields we can form the constants of motion $Q_I=\xi_{(I)}^\alpha\pi_\alpha$, where $\pi_\alpha=\partial_{x'^\alpha}L$ are the momenta, along with the constant $Q_\eta=\eta^\alpha\pi_\alpha+\int N du$ and calculate the functions $w(u),f(u)$. In order to simplify the results we can switch to the time variable $\tau$ with $\drm \tau=N(u)\,\drm u$. Denoting with $\kappa_I$ the three constants of motion which correspond to the Killing fields and with $k_h$ the constant arising from the homothetic filed, we have
\bsub\label{eqk}
\bal
32\exp \frac{\left(1+2\sqrt{3\lambda}\right)\left( \sqrt{\lambda}f(\tau)+w(\tau) \right)}{2\sqrt{\lambda}}  \left( \sqrt{\lambda}f'(\tau)+w'(\tau) \right)&=\kappa_1\\
32\exp \frac{\left(1-2\sqrt{3\lambda}\right)\left( \sqrt{\lambda}f(\tau)-w(\tau) \right)}{2\sqrt{\lambda}}  \left( \sqrt{\lambda}f'(\tau)-w'(\tau) \right)&=\kappa_2\\
32\exp\left( f(\tau)+2\sqrt{3}w(\tau) \right) \left( 2\sqrt{3}\lambda f'(\tau)+w'(\tau) \right)&=\kappa_3\\
\frac{32}{\sqrt{3}}\exp\left( f(\tau)+2\sqrt{3}w(\tau) \right)w'(\tau)&=\tau-k_h,
\eal
\esub

The above four equations can be solved \emph{algebraically} for the functions $f(\tau),w(\tau)$ and their derivatives $f'(\tau),w'(\tau)$; but after that, we must demand validity of the consistency equations $f'(\tau)=\drm f(\tau)/\drm \tau, w'(\tau)=\drm w(\tau)/\drm \tau$.

From the form of the equations \eqref{eqk}, it is obvious that we have to consider two cases, where the constant $\lambda$ equals $\frac{1}{12}$ or not.

$\bullet$ {\bf Case I: $\lambda=\dfrac{1}{12}$}.

The consistency equations imply the following relations between the constants $\kappa_i$
\bal
\kappa_3=\kappa_1,\, \kappa_2=-\frac{32}{\kappa_1},
\eal
so the functions $f(\tau),\,w(\tau)$ are given by
\bsub
\bal
f(\tau)&=c_1-\frac{\sqrt{3}}{\kappa_1}\,\tau+\frac{1}{2}\ln\left( 2\sqrt{3}\tau-k \right)\\
w(\tau)&=-\frac{\sqrt{3}}{6}\left( c_1 +\ln\frac{32}{\kappa_1} \right)+\frac{1}{2\kappa_1}\,\tau+\frac{\sqrt{3}}{12}\ln\left( 2\sqrt{3}\tau-k \right),
\eal
\esub
where $k=2\sqrt{3}c_h+\kappa_1$. The values of the original functions $a(\tau), V(\tau), F(\tau)$ and $\phi(\tau)$ can be deduced from the parametrization \eqref{FVpar} and \eqref{trans}, i.e.
\bsub\label{fcaseI}
\bal
a(\tau)&=\frac{(2\kappa_1)^{1/6}}{h(\tau)}\left( 2\sqrt{3}\tau-k\right)^{1/12}\exp\left(\frac{\sqrt{3}}{6\kappa_1}\tau-\frac{c_1}{6}\right)   \\
V(\tau)&=\left( 2\sqrt{3}\tau-k\right)^{1/2}h^4(\tau)\exp\left( c_1-\frac{\sqrt{3}}{\kappa_1}\,\tau\right)    \\ \label{fcaseIc}
F(\tau)&=\frac{1}{4}h^2(\tau)  \\ \label{fcaseId}
\phi'(\tau)^2&=\frac{3}{\epsilon}\,h'^2(\tau)- \frac{1}{\epsilon} \left(\frac{2\sqrt{3}\tau-k-\kappa_1}{2\kappa_1\left( 2\sqrt{3}\tau-k \right)} \right)^2 h^2(\tau).
\eal
\esub

Thus we have an \emph{infinite} number of coupling functions $F(\phi)$ and interacting potentials $V(\phi)$ resulting from the \emph{infinite} choices of the arbitrary function $h(\tau)$.

If the actual form $F(\phi)$ is needed, it can be derived as follows: choose a function $h(\tau)$, calculate the functional form of $\phi(\tau)$ from \eqref{fcaseId}, take the inverse of that function in order to get $\tau=r(\phi)$ and then substitute the result in \eqref{fcaseId}.

As an example let $\epsilon=1,k=0,\kappa_1=2\sqrt{3},\, h(\tau)=\sqrt{48/143}\,e^\tau/\tau$, then $\phi(\tau)=c\pm e^\tau/\tau$ (where the plus sign emerges when $\tau>1$ while the minus sign when $\tau<1$), then $h=\pm\sqrt{48/143}\left( \phi -c\right)$ and finally $F(\phi)=12/143(\phi-c)^2$ along with $V(\phi)=c_2\left( \phi - c\right)^{7/2}$.

As it is common in General Relativity, the constants that are appearing in the solution set, are not all essential, i.e. they can be eliminated by a proper redefinition of them, along with a coordinate transformation. In our case the redefinitions $k=2\sqrt{3}\gamma,\,k_1=1/(\sqrt{3}\alpha),\, \exp c_1=\alpha/(4\sqrt{2\sqrt{3}}\beta)$ and the transformation $r \mapsto e^{-c_1/3}\alpha^{2/3}/(2\sqrt{2}3^{1/12})\, r$, bring the solution space into the form

\bsub\label{final_caseI}
\bal
a(\tau)&=\frac{\beta}{h(\tau)}\, e^{\alpha\tau/2}\left( \tau-\gamma \right)^{1/12}   \label{a_I}\\
V(\tau)&=\frac{\alpha\,h^4(\tau)}{4\beta^2}\, e^{-3\alpha\tau}\sqrt{\tau-\gamma}    \label{V_I}\\
F(\tau)&=\frac{1}{4}h^2(\tau)  \label{F_I}\\
\phi'(\tau)^2&=\frac{3}{\epsilon}\,h'^2(\tau)- \frac{1}{48\epsilon} \left(\frac{6\alpha\tau-6\alpha\gamma-1}{\tau-\gamma} \right)^2 h^2(\tau), \label{psi_I}
\eal
\esub
yielding the line element
\bal\label{ds_caseI}
\drm s^2=\frac{\beta^2\, e^{3\,\alpha\,\tau}}{h^2(\tau)\sqrt{\tau-\gamma}} \left( -\drm \tau^2+e^{-2\,\alpha\,\tau}\left( \tau-\gamma \right)^{5/3} \left( \drm r^2 + r^2\drm \theta^2+r^2 \sin^2\theta\, \drm \varphi^2 \right) \right),
\eal
with $\tau>\gamma$.

$\bullet$ {\bf Case II: $\lambda \neq \dfrac{1}{12}$}.

In this case the consistency equations imply only one relation for the constants $\kappa_i$
\bal
\kappa_2=-\frac{32}{\kappa_1},
\eal
yielding the functions $f(\tau),\,w(\tau)$
\bsub
\bal
f(\tau)&=\frac{1}{s+1}\ln\frac{3k_1(s+1)(\tau-\beta)}{32s}-\frac{1}{s-1}\ln\frac{(s-1)(\tau+\alpha)}{k_1s}\\
w(\tau)&=\frac{s}{2\sqrt{3}}\left( \frac{1}{s+1}\ln\frac{3k_1(s+1)(\tau-\beta)}{32s}+\frac{1}{s-1}\ln\frac{(s-1)(\tau+\alpha)}{k_1s} \right),
\eal
\esub
where the redefinitions of the various constants are $\lambda=s^2/12,\,\kappa_1=\sqrt{3}\,k_1$ and $c_h=(\alpha(1-s)+\beta(1+s))/(2s), \kappa_3=\sqrt{3}(s^2-1)(\alpha+\beta)/(2s)$. Once more the values of the original functions $a(\tau), V(\tau), F(\tau)$ and $\phi(\tau)$ can be deduced from the parametrization \eqref{FVpar}, i.e.
\bsub\label{fcaseII}
\bal
a(\tau)&=\frac{2}{h(\tau)} \left( \frac{3k_1(s+1)(\tau-\beta)}{32s} \right)^{s/6(1+s)} \left( \frac{(s-1)(\tau+\alpha)}{k_1s} \right)^{s/6(1-s)}\\
V(\tau)&= \left( \frac{3k_1(s+1)(\tau-\beta)}{32s} \right)^{1+s} \left( \frac{(s-1)(\tau+\alpha)}{k_1s} \right)^{1-s}h^4(\tau)\\
F(\tau)&=\frac{1}{4}h^2(\tau) \\
\phi'(\tau)^2&=\frac{3}{\epsilon}\,h'^2(\tau)- \frac{s^2\left(2\tau-(s-1)\alpha-(s+1)\beta \right)^2}{12\epsilon (s^2-1)^2(\tau-\beta)^2(\tau+\alpha)^2}h^2(\tau).
\eal
\esub

Exactly as in case I, we have an \emph{infinite} number of coupling functions $F(\phi)$ and interacting potentials $V(\phi)$ arising from the appearance of the arbitrary function $h(\tau)$.

The following redefinition and the transformation of $r$--coordinate
\bal
k_1&=2^{-(s-1)(s-4)/(2s)}3^{-(s+2)(s-1)/(2s)}\non\\
&(s-1)^{-(s+1)(s-2)/(2s)}(s+1)^{-(s-1)(s+2)/(2s)}\gamma^{-(s^2-1)/s}\non\\
r &\mapsto  2^{(3-2s)/(3+3s)}3^{-(3+2s)/(3+3s)}k_1^{(2s)/(3-3s^2)}\non\\
&(s-1)^{(3-2s)/(-3+3s)}(s+1)^{-(3+2s)/(3+3s)}s^{(6-4s^2)/(3-3s^2)}\,r\non
\eal
considerably simplifies the form of the solution space
\bsub\label{final_caseII}
\bal
a(\tau)&=\frac{\gamma}{h(\tau)}\left( \tau +\alpha\right)^{s/6(s-1)} \left( \tau -\beta\right)^{s/6(s+1)} \label{a_II}\\
V(\tau)&=\frac{s^2\,h^4(\tau)}{6\gamma^2\left(s^2-1\right)}\left( \tau +\alpha\right)^{1/(1-s)} \left( \tau -\beta\right)^{1/(s+1)}   \label{V_II}\\
F(\tau)&=\frac{1}{4}h^2(\tau)  \label{F_II}\\
\phi'(\tau)^2&=\frac{3}{\epsilon}\,h'^2(\tau)- \frac{s^2\left(2\tau-(s-1)\alpha-(s+1)\beta \right)^2}{12\epsilon (s^2-1)^2(\tau-\beta)^2(\tau+\alpha)^2}h^2(\tau). \label{psi_II}
\eal
\esub
yielding the line element
\bal\label{ds_caseII}
\drm s^2=\frac{\gamma^2\left( \tau +\alpha\right)^{(2-s)/(s-1)}}{h^2(\tau)\left( \tau -\beta\right)^{(s+2)/(s+1)}} \left( -\drm \tau^2+\frac{\left( \tau +\alpha\right)^{n}}{\left( \tau -\beta\right)^{m}} \left( \drm r^2 + r^2\drm \theta^2+r^2 \sin^2\theta\, \drm \varphi^2 \right) \right),
\eal
where $n=\dfrac{4s-6}{3s-3},\,m=-\dfrac{4s+6}{3s+3}$.

\subsubsection{Mini-superspace with lesser autonomous integrals of motion}
In this subsection we investigate what the result of the previous investigation would be if the assumption of maximal symmetry for $G_{\alpha\beta}$ was relaxed, i.e. if we demanded less than three autonomous integrals of motion. As it is well known, in two dimensions the general metric can be brought in a conformally flat form. We thus, need to investigate the case where the conformal factor is such that the mini--supermetric \eqref{scGdd} is not flat. In order to find its Killing/homothetic fields we first begin by enumerating all the possibilities. The maximum number of Killing fields for an $n$--dimensional metric is $n(n+1)/2$ thus in our case this number equals three.
\begin{itemize}
\item If the metric admits three Killing fields, then its either flat or maximally symmetric; the first possibility is already checked, while the second (as we have already proved) is not admissible.
\item If the metric admits two Killing fields $\xi_{(1)},\xi_{(2)}$, then the possible Lie algebras these fields can span, are either the Abelians $2A_1=\langle\partial_x,\partial_y\rangle$ and $2A_1=\langle\partial_x,y\partial_x\rangle$, or the non-Abelians $A_2=\langle\partial_x,e^x\,\partial_y\rangle$ and $A_2=\langle\partial_x,x\,\partial_x\rangle$, see e.g. \cite{Popovych:RealLieAlg,Terzis:2013Faith}.

In the Abelian case the second algebra yields a degenerate metric, while the first algebra reproduces a flat metric, since $G_{\alpha\beta}=const.$ (and of course admits a third Killing field).

In the non--Abelian case  the second algebra yields a degenerate metric, while the first algebra reproduces a metric with a constant Ricci scalar, i.e. a maximally symmetric metric.
\end{itemize}

Finally the only case which is left to discuss is when the scaled supermetric \eqref{scGdd} admits only one Killing field. First of all, let us state some general facts; let $h_{\alpha\beta}=h_{\alpha\beta}(x,y)$ a two dimensional metric which admits a Killing field $\xi^\alpha$, then it is always possible to bring it into its normal form, i.e. $\xi=\partial_y$. As a result the metric can be put in the special conformal form
\bal \label{metonev}
h_{\alpha\beta}=\Omega(x)
\begin{pmatrix}
1 & 0\\ 0 & \epsilon
\end{pmatrix},
\eal
see Appendix \ref{2Dcase}. Obviously with the help of the transformation $x\to y,\, y\to x$ we can make the conformal factor $\Omega$ a function of $x$.

Let us now return to the supermetric \eqref{scGdd} and apply once more the transformation $a=\exp(w/\sqrt{3})/\sqrt{F(\phi)}$, which turns the line element into the form
\bal\label{confGdd}
\drm s^2=-\frac{4e^{2\sqrt{3}w}V(\phi)}{F^2(\phi)}\left( \drm w^2+\frac{\epsilon\,F(\phi)-3\,F'^2(\phi)}{4F^2(\phi)}\, \drm\phi^2  \right).
\eal
Employing the transformation $\phi=r(y)$ such that
\bal
\sqrt{ \left| \frac{\epsilon\,F(\phi)-3\,F'^2(\phi)}{4F^2(\phi)} \right|}\,\drm\phi=\drm y,
\eal
we bring the metric \eqref{confGdd} into the desired form
\bal \label{metwy}
\drm s^2=-\frac{4e^{2\sqrt{3}w}V(r(y))}{F^2(r(y))}\left( \drm w^2+\epsilon\,\drm y^2  \right).
\eal

In order for this line element to admit one Killing field, there must exist a transformation that brings \eqref{metwy} into the form \eqref{metonev}. When $\frac{V}{F^2} = c e^{\mu y}$ the space is flat and thus admits three Killing fields, this case has the space of solutions described by the sets \eqref{final_caseI} and \eqref{final_caseII}. For all other functional forms of $\frac{V}{F^2}$, the space is not flat and cannot be transformed into a form analogous to \eqref{metonev}.

\section{Physical interpretation and physical parameters}

In a FLRW universe some of the physical observation parameters, are the Hubble parameter $H$ and the dimensionless parameters deceleration $q$ and jerk $j$, see for example \cite{Blandford:2004,Nair:2011}. The Hubble parameter quantifies the expansion of the universe; the deceleration parameter nowadays measures the acceleration of the universe whereas the jerk parameter is needed since the universe was once decelerating and is now accelerating.

Their definitions in comoving coordinates ($\drm s^2=-\drm t^2+a^2(t)\drm r^2+a^2(t)r^2\drm \Omega^2$) are given by
\bal
H=\frac{a'(t)}{a(t)}, \quad q=-\frac{a(t)\,a''(t)}{a'^2(t)},  \quad j=\frac{a^2(t)\,a'''(t)}{a'^3(t)}.
\eal

The solution sets \eqref{final_caseI} and \eqref{final_caseII}, are not referring to comoving coordinates, but it is an easy task to make the transition, from the time coordinate $\tau$ to the desired one $t$. If we follow the redefinition of the lapse function $N(u)$ and the time re--parameterization $\drm \tau=N(u)\drm u$, we then see that these solutions, are expressed as the line element
\bal\label{ds_tau}
\drm s^2=-u(\tau)^2 \drm \tau^2 +a^2(\tau)\left( \drm r^2+r^2\,\drm\Omega^2 \right),
\eal
thus the two coordinates are connected by
\bal
u(\tau) \drm \tau=\drm t,
\eal
and the time derivatives of the scale factor $a(t)$ are
\bal
a'(t)=\frac{\drm a(\tau)}{u(\tau)\drm \tau},\quad a''(t)=\frac{\drm}{u(\tau) \drm \tau}\left( \frac{\drm a(\tau)}{u(\tau) \drm \tau}\right), \quad \dots
\eal

For each one of the two solutions the aforementioned parameters are quite cumbersome, due to the existence of the arbitrary function $h(\tau)$, but are quite straightforward to be calculated. We only present the form of the Hubble parameter for each case
\bsub
\bal
H_{I}&=\frac{1}{12\beta\,\left( \tau-\gamma \right)^{1/4}} e^{-3\alpha\tau/2}\left(  -12\left( \tau-\gamma\right)h'(\tau) +\left( 6\alpha\tau-6\alpha\gamma+1 \right) h(\tau)\right)  \\
H_{II}&=\frac{\left(\tau+\alpha \right)^{s/(2-2s)}\left(\tau-\beta \right)^{-s/(2+2s)}}{6\gamma\,\left(s^2-1 \right)}\Big( -6\left(s^2-1\right)\left(\tau-\beta\right) \left(\tau+\alpha\right) h'(\tau)+\non\\
&\ph{=} s\left( \left(s-1\right)\alpha-\left(s+1\right)\beta+2s\tau \right) h(\tau) \Big).
\eal
\esub

As we have mentioned in the Introduction, the duality of scalar field/fluid is widely used in cosmology, thus we are going to apply this procedure in our case, for details see \cite{Madsen1, Madsen2, Pimentel}.

The equations of motion \eqref{EoM1} can be rewritten as
\bal
R_{ij}-\frac{1}{2}Rg_{ij}=\frac{1}{F(\phi)}\left(T_{ij} +\nabla_i\nabla_j F(\phi)+g_{ij}\Box F(\phi)\right)\Rightarrow E_{ij}=T^{(\phi)}_{ij},
\eal
where $E_{ij}$ is the Einstein tensor and $T^{(\phi)}_{ij}$ is the effective energy--momentum associated with the
scalar field. With this energy--momentum tensor we want to associate an energy--momentum tensor of an imperfect fluid
\bal
T^{(imf)}_{ij}=\left( \rho+p \right)u_iu_j+p g_{ij}+2q_{(i}u_{j)}+\pi_{ij},
\eal
where $\rho$ is the energy density of the fluid, $u_i$ the 4--velocity, $q_i$ the heat flux vector, $p$
the pressure and $\pi_{ij}$ the anisotropic stress tensor. The relations that make the identification possible are
\bsub
\begin{alignat}{2}
\Pi_{mn}&=T^{(\phi)}_{ij} h^i{}_m h^j{}_n=p h_{mn}+\pi_{mn}  &\quad\quad \pi_{mn}&=\Pi_{mn}-\frac{1}{3}\Pi_k{}^k h_{mn}=\Pi_{mn}-p h_{mn}\\
\rho&=T^{(\phi)}_{ij}u^i u^j & p&=\frac{1}{3}\Pi_i{}^i\\
q_k&=-T^{(\phi)}_{ij}u^i h^j{}_k,
\end{alignat}
\esub
 where $h_{ij}$ is the projection tensor orthogonal to velocity $u_i$ defined by
\bal
h_{ij} = g_{ij}+ u_i u_j \quad \text{with} \quad u_i u^i=-1.
\eal

The natural choice of the 4--velocity $u_i$ is the one which is associated with the  normalized derivative of the scalar field $\phi$, i.e.
\bal
u_i=\frac{\nabla_i\phi}{\sqrt{-\nabla_j\phi \nabla^j\phi}},
\eal
where we have assumed that $\nabla_i\phi$ is timelike in order to describe a physical fluid. Furthermore the kinematical quantities of the fluid that are of interest, and appear in the decomposition of the covariant derivative of the velocity \cite{ellis2012relativistic}
\bal
\nabla_i u_j=-\dot{u}_i u_j+\omega_{ij}+\sigma_{ij}+\frac{1}{3}\theta h_{ij},
\eal
are
\bal
\dot{u}_i=u^j\nabla_j u_i, \quad \theta=\nabla_i u^i, \quad \sigma_{ij}=\nabla_{(i} u_{j)}+\dot{u}_{(i}u_{j)}-\frac{1}{3}\theta h_{ij}, \quad \omega_{ij}=\nabla_{[i} u_{j]}+\dot{u}_{[i}u_{j]}
\eal
i.e. the acceleration, the expansion, the shear and the rotation of the fluid respectively.

It is quite remarkable that in our case, \emph{irrespectively} of the solution space the heat flow $q_i$ along with the anisotropic stress tensor $\pi_{ij}$ are zero, thus the stress tensor $T_{ij}^{(\phi)}$ mimics
a perfect fluid. Moreover the fluid has zero acceleration, is shear free, exhibits no rotation and the expansion is three times the Hubble parameter with an opposite sign. The forms of the pressure $p$ and the energy density $\rho$ for the case I are
\bsub
\bal
p&=\frac{e^{-3\alpha\tau}}{48\beta^2\sqrt{\tau-\gamma}}\Bigg(96\left(\tau-\gamma\right)^2 h(\tau)h''(\tau)- 144\left(\tau-\gamma\right)^2h'^2(\tau)\non\\
&-8\left(\tau-\gamma \right)\left(6\left(\tau-\gamma\right)\alpha-11 \right)h(\tau) h'(\tau)\non\\
&+\left(36\alpha^2\left(\tau-\gamma\right)^2-36\alpha (\tau-\gamma)+1 \right) h^2(\tau)\Bigg) \\
\rho&=\frac{e^{-3\alpha\tau}}{48\beta^2\sqrt{\tau-\gamma}}\Big(12\left(\tau-\gamma\right) h'(\tau)-
\left(6\alpha\tau-6\alpha\gamma +1\right) h(\tau)\Big)^2,
\eal
\esub
while for the case II are given by
\bsub
\bal
p&=d(\tau) \Bigg(6\left(s^2-1\right)(\tau+\alpha)(\tau-\beta) h(\tau)h''(\tau)-9\left(s^2-1\right)(\tau+\alpha)(\tau-\beta)h'^2(\tau)\non\\
&+\left( (5s+6)(s-1)\alpha-(5s-6)(s+1)\beta+2\left( 5s^2-6\right)\tau   \right) h(\tau)h'(\tau)\non\\
&-\frac{s^2}{4}\left(4  -\frac{\left( 2\tau-(s+1)\beta-(s-1)\alpha\right)^2}{\left(s^2-1\right)(\tau+\alpha)(\tau-\beta)} \right) h^2(\tau) \Bigg) \\
\rho&=d(\tau) \Bigg(9\left(s^2-1\right)(\tau+\alpha)(\tau-\beta) h'^2(\tau)-3s\left(2s\tau+(s-1)\alpha-(s+1)\beta\right)h(\tau) h'(\tau)\non\\
&+\frac{s^2\left( 2s\tau+(s-1)\alpha-(s+1)\beta \right)^2}{4\left(s^2-1\right)(\tau+\alpha)(\tau-\beta)}  h^2(\tau)\Bigg),
\eal
\esub
where
\bal
d(\tau)=\frac{(\tau+\alpha)^{1/(1-s)}(\tau-\beta)^{1/(s+1)}}{3\gamma^2\left(s^2-1\right)} . \non
\eal

The above procedure of evaluating the pressure $p$ and the energy density $\rho$ has already been criticized as being non--physical in the case of vacuum scalar--tensor theory. In \cite{RomeroBarros} the authors started from the solutions of the vacuum field equations of Brans-Dicke scalar-tensor theory of gravity and calculated the corresponding energy--momentum tensor for the perfect fluid. Their conclusion was that \emph{''The examples presented in this paper seems to
suggest that this sort of equivalence is sometimes purely formal and rather artificial.''}. Thus in order for one to be on the safe side he must demand a physical character of the presented values of the pressure $p$ and the energy density $\rho$. The minimum assertions that can guarantee that sort of physical acceptance is the various energy conditions, which for the perfect fluid can then be formulated in terms of the eigenvalues of this energy momentum tensor:
\begin{itemize}
\item The weak energy condition stipulates that $\rho \ge 0, \; \; \rho + p \ge 0 .$

\item The null energy condition stipulates that $\rho + p \ge 0 $.

\item The strong energy condition stipulates that $\rho + p \ge 0, \; \; \rho + 3 p \ge 0 $.

\item The dominant energy condition stipulates that $\rho \ge |p| .$
\end{itemize}

Below we present a solution that satisfies the above energy conditions, expect the strong energy one. The validation of the strong energy condition is being criticized nowadays; one of its violation can be seen from the recent observational data regarding the acceleration of our universe, for more details see \cite{Visser:1999}.

\section{Special solutions}

In this section we intend to explore some physical consequences of the solution space found. To this end, we select some particular forms for the free function $h(\tau)$ parameterizing the different solutions. We thus arrive at the three cases given below.

\subsection{Energy Complete Solution}
 As we have already mentioned each solution space \eqref{final_caseI}, \eqref{final_caseII} is modeled by the existence of the arbitrary function $h(\tau)$. A natural choice would be to make the $00$ component of the line element \eqref{ds_tau} equal to minus one, thus bringing it to comoving coordinates; it is to be noticed that with this procedure we \emph{do not} apply any coordinate transformation but we only make a specific choice of $h(\tau)$. For case I the function $h(\tau)$ which accomplishes this is
\bal\label{h_caseI}
h(t)=\beta\,e^{3\alpha\, t/2}\left( t-\gamma \right)^{-3/4},
 \eal
(for simplicity we write $t$ instead of $\tau$) and the line element \eqref{ds_tau} reads
\bal\label{ds_t}
\drm s^2=-\drm t^2+e^{-2\alpha t}t^{5/3}\left( \drm r^2+r^2\drm \theta^2 +r^2\sin^2\theta\drm \phi^2\right).
\eal

For the above line element the Hubble parameter $H$, the deceleration parameter $q$ and the jerk parameter $j$ are
\bal\label{Hqj}
H=\frac{5}{6t}-\alpha,\quad q=-1+\frac{30}{\left(  6\alpha t-5 \right)^2}\, \quad j=1+90\,\frac{ 6\alpha t-1 }{\left(  6\alpha t-5 \right)^3}.
\eal

Furthermore the pressure and the energy density of the perfect fluid are
\bal
p=\frac{20-\left( 6\alpha t-5\right)^2}{12t^2}, \quad \rho=\frac{\left( 6\alpha t-5\right)^2}{12t^2},
\eal
 which yield an equation of state $p=w\rho$, with variable equation of state parameter $w$
\bal
w=\frac{20-\left( 6\alpha t-5\right)^2}{\left( 6\alpha t-5\right)^2}\Rightarrow w=-1+\frac{20}{\left( 6\alpha t-5\right)^2}.
 \eal

\begin{figure}[t]
\centering
\begin{minipage}{0.45\linewidth}
\includegraphics[width=0.9\textwidth]{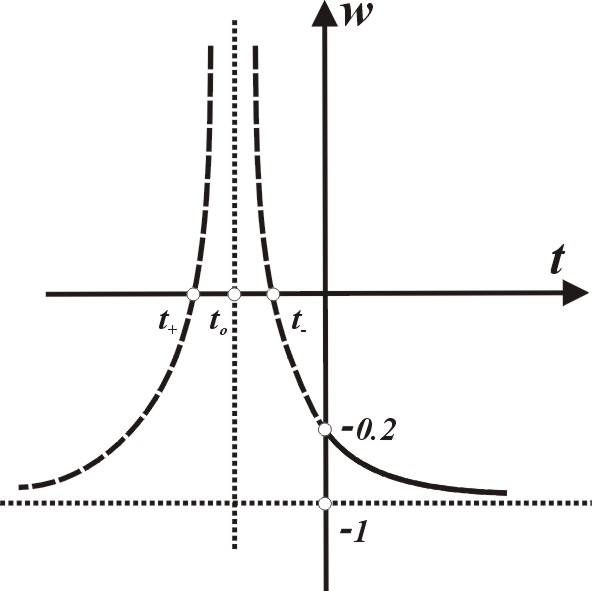}
\caption{The zeros of the state parameter $w$ occur at  $t_\pm=\left(5\pm2\sqrt{5} \right)/(6\alpha)$.}
\label{fig:w-t}
\end{minipage}
\,
\begin{minipage}{0.45\linewidth}
\includegraphics[width=0.9\textwidth]{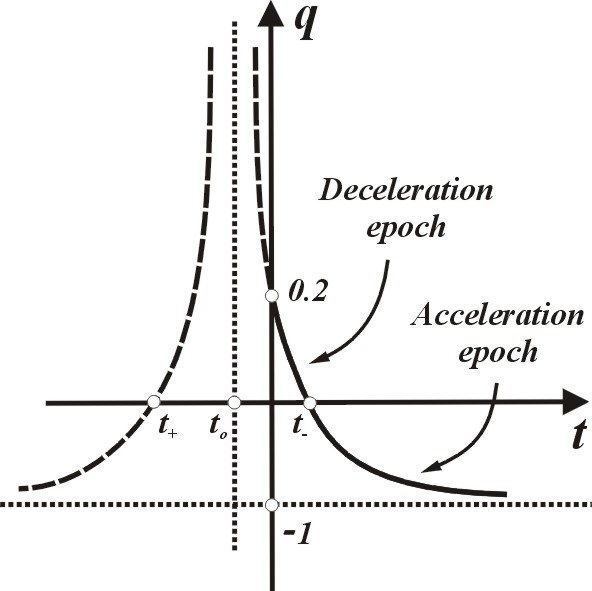}
\caption{The deceleration epoch $q>0$ is followed by an acceleration one $q<0$.}
\label{fig:q-t}
\end{minipage}
\end{figure}

From the above we conclude the following facts
\begin{enumerate}
\item It is easy to see that the energy conditions, (expect the strong energy one) are satisfied for any value of the parameter $\alpha$ and for $t \neq 0$. Thus the induced perfect fluid is a physical one, although the parameter $w$ is $t$--dependent.
\item For $t=0$ the parameter $w$ equals $-1/5$ for every value of the constant $\alpha$, while for $t \to \infty, w\to -1$, i.e. the fluid behaves like a cosmological constant see figure \eqref{fig:w-t}.
\item In order to have a universe that is expanding we must require that the constant $\alpha$ in line element \eqref{ds_t} to be negative, $\alpha<0.$
\item The deceleration parameter $q$, equals $q=1/5$ at $t=0$ and changes its sign at $t_-=\disfrac{n-\sqrt{n}}{\alpha},\, n=5/6$; thus we are describing a universe that initially is expanding though decelerating, and after $t_-$ the expansion is accelerating, a fact that is supported by recent observations see \cite{Riess:2004dec}. This behavior is shown in figure \ref{fig:q-t}.
\item The jerk parameter is always positive for $t>0$ and for all $\alpha$, thus the acceleration is always increasing.
\item Since both the pressure $p$ and the energy density $\rho$ are functions of $t$ we can find a relation between the two of them; the result is $p=\dfrac{12}{5}\left(\alpha\pm \sqrt{\dfrac{\rho}{3}} \right)^2-\rho$.
\end{enumerate}

We can use the above results to estimate the age of the universe $t_0$. From the observations of the Hubble Space Telescope Key project \cite{Freedman:2000cf}, the present Hubble parameter is constrained to be
\bal
H_0^{-1}= 9.776h^{-1}\, Gyr, \quad 0.64 < h < 0.80,
\eal
where the subscript $0$ indicates present time. We could insert this value into \eqref{Hqj} to estimate $t_0$, but we must know the value of the constant $\alpha$. In order to calculate $\alpha$ we will make use of the knowledge of the redshift $z_{cr}$ where the cosmic jerk happened (at $t_-$), i.e. the redshift where the deceleration parameter changed sign indicating that the current epoch of cosmic acceleration was preceded by a cosmic deceleration one. The Supernova Search Team constrained this value to $z_{cr}=0.46 \pm 0.13$  \cite{Riess:2004dec}. From the relation of the scale factor $a(t)$ and the redshift $a_0/a=1+z$ we
have
\bal\non
t&=t_0: a_0=e^{-\alpha t_0}t_0^{5/6}\\
t&=t_-: a_0=e^{-\alpha t_-}t_-^{5/6}(1+z_{cr})
\non,
\eal
and using \eqref{Hqj} $\alpha=5/(6t_0)-H_0$ we end up
\bal
e^{-\sqrt{n}+x}=\left( \frac{\sqrt{n}-n}{x-n} \right)^n(1+z_{cr}), \quad n=\frac{5}{6},\quad x=H_0\,t_0.
\eal

Using $z_{cr}=0.46,\, h=0.72$ we have $x=0.952737$ thus $t_0=12.9361\, Gyr$ a very fine result for this model, since the most recent WMAP3 data
produces a value of $t_0 = 13.73^{+0.13}_{-0.17}\, Gyrs$ (assuming an \textrm{$\Lambda$}CMD model) \cite{Spergel:2006hy}.

\subsection{Singular supermetric - A dust solution}
The previous analysis of the solution space is valid in the case where the mini-supermetric $G_{\alpha\beta}$ is not singular, i.e. $\det{G_{\alpha\beta}} \neq 0$. Thus we have to consider separately the case where  $\det{G_{\alpha\beta}} = 0$.

The determinant of the scaled supermetric \eqref{scGdd} is
\bal
G=12a^{10}V^2(\phi) \left( \epsilon\, F(\phi)-3F'^2(\phi) \right)   \non
\eal
which is zero when
\bal
F(\phi)=\frac{\epsilon}{12}\left( \phi - c\right)^2.
\eal
Taking for simplicity $\epsilon=1$ and $N=1$, we can calculate $V(\phi)$ from the $00$ component of the field equations \eqref{EoM1}
\bal
V(\phi)=\frac{2}{a^4\left( \left(\phi-c \right) a'+a\, \phi'\right)^2} \non.
\eal
Substituting the above $V(\phi)$ in the field equation \eqref{EoM2}, we can solve for $\phi''$ with the help of which all the components of \eqref{EoM1} are made proportional to
\bal
4\alpha \phi'+6\left( \phi-c\right) a' =0, \non
\eal
which can be integrated to
\bal
\phi=c+c_1\,a^{-3/2},
\eal
where $c_1$ is a constant of integration. With the above information at hand, equation \eqref{EoM2} reads
\bal
2a a''-5a'^2=0\Rightarrow a=\frac{c_2}{\left( 3t+2c_3\right)^{2/3}},
\eal
where $c_2,c_3$ are constants of integration. The coordinate transformation $t\to -2c_3/3+c_1^6/(648\, c_2^3\, t)$ along with the redefinition $c_2 \to \left(c_1^{4/3}6^{-2/3}\right)\kappa^{1/3}$, makes $\kappa$ a multiplicative constant, and simultaneously brings the line element to the form
\bal\label{ds_p=0}
\drm s^2=-\drm t^2+t^{4/3}\left( \drm r^2+r^2 \drm \theta^2+r^2\sin^2\theta \drm \phi^2\right).
\eal

The overall constant $\kappa$ does not appear in the line element, since it admits the homothetic vector field $h=3t\partial_t+r\partial_r$ which can absorb it. The final form of the scalar field $\phi(t)$, the potential $V(t)$, and the function $F(t)$ are
\bal
\phi(t)=c+\frac{c_1}{\kappa^{3/2}t},\quad V(t)=\frac{c_1^2}{18\kappa^5 t^4},\quad F(t)=\frac{c_1^2}{12\kappa^3 t^2}.
\eal

We can calculate the values of the Hubble, the deceleration and the jerk parameters for line element \eqref{ds_p=0}
\bal
H=\frac{2}{3t}, \quad q=\frac{1}{2}, \quad j=1,
\eal
thus we are describing a universe that expands while decelerating. Furthermore if we apply the identification of the scalar field to the perfect fluid we find that $p=0$ and $\rho=4/(3\kappa^2 t^2)$; thus we have the dust solution of Friedmann \cite{Friedmann1924, Friedmann1924en}.

It is quite interesting that this General Relativity solution is found from the perspective of scalar--tensor gravity, as an exceptional case.

\subsection{A cosmological solution with a constant parameter of state $w$}
Solution sets \eqref{final_caseI} and \eqref{final_caseII} can be reduced into General Relativity's theme by demanding the constancy of the function $F(\phi)$. A wide class of cosmologies can be inferred from these solutions; we are going to present one with constant parameter of state $w$, i.e. $p=const. \, \rho$.

For the set \eqref{final_caseII} in order to have $F(\phi)=1$ we must take $h(\tau)=2$. If we choose $\beta=-\alpha$ it is easy to see that the parameter of state reads
\bal
w=-1+\frac{2}{s^2}.
\eal

Performing the coordinate transformation $\tau=z^{z/(z-2)t^z-\alpha},\, z=s^2-1$ and making the redefinition $\gamma=2z^{s^2/(s^2-3)}\kappa$, in order to make $\kappa$ an overall constant, we end up with the line element
\bal
\drm s^2=-\drm t^2+t^{2s^2/3} \left( \drm r^2+r^2 \drm \theta^2+r^2\sin^2\theta\, \drm\phi^2  \right),
\eal
where $t>0$. This line element admits the homothetic vector field $h=3t\partial_t+(3-s^2)r\partial_r$ which justifies the omission of $\kappa$ in front of it. The pressure $p$ and the energy density $\rho$ of the perfect fluid are
\bal
p=\frac{s^2\left( -s^2+2  \right)}{3\,t^2},\, \rho=\frac{s^4}{3\,t^2} \Rightarrow w=-1+\frac{2}{s^2}.
\eal
This solution can be found in \cite{Stephani2003exact}, p.212 equation (14.8b), and it describes a decelerating expanding universe with constant deceleration parameter, since
\bal
H=\frac{2\,s^2}{3\,t}, \quad q=-1+\frac{3}{2\,s^2}, \quad j=1+\frac{9}{2\,s^4}(1-s^2).
\eal

\section{Discussion}

The use of symmetries in the process of acquiring new solutions is widespread in mathematical cosmology. In our case this approach was applied, in the context of a spatially flat FLRW space--time, to the mini--superspace Lagrangian of the scalar tensor theory of gravity. In order to acquire all the existing autonomous, linear in the momenta integrals of motion for a given cosmological system, its singular nature must necessarily be taken into account \cite{CDT2014,tchris_sch}. One of the main aims of the present work is to highlight exactly this: the immense possibilities that can be explored by taking into account the reparametrization invariance of constrained systems.

Unfortunately, a common practise in the literature is, to gauge fix the lapse function (usually to $1$), so that the theory of Noether symmetries for regular systems can be applied. However, this process is misleading in what regards the properties of the system under consideration. As it is known, the mini--superspace Lagrangians ensuing from cosmological systems are singular and belong to the general form \eqref{Lag}. By gauge fixing the lapse, for example $N=1$, the new fixed Lagrangian reads
\begin{equation}
L_{\mathrm{fixed}}=\frac{1}{2} G_{\alpha\beta} x'^\alpha\, x'^\beta - U(x)
\end{equation}
and describes a system different from \eqref{Lag}. Of course the former can admit the same solution if one uses the constraint equation $\frac{\partial L}{\partial N}=0$ of the initial system, as an \emph{ad hoc} condition. Nevertheless, as far as the search of symmetries is concerned, this procedure becomes too restrictive. The fixing of the lapse annihilates the freedom of the reparametrization invariance that in itself, as shown in \cite{CDT2014} and \cite{tchris_sch}, is a source for the emergence of linear in the momenta integrals of motion which are not obtained in the theory of regular systems.

All the previous arguments can be made clearer in the context of scalar tensor gravity that we have treated in this paper. By comparing results with \cite{tsamp}, where the authors start from the same action \eqref{action}, but in the process follow the gauge fixing approach, one can see that: Under the same condition that we used here, i.e. the mini-superspace being maximally symmetric, they are led to a specific functional form for $F(\phi)$, say $F_{\mathrm{fixed}}(\phi)$, for which $G_{\alpha\beta}$ is flat. Subsequently, they apply each of the three killing fields to the potential $U(x)$ acquiring three different scalar field potentials. Each of them is used to describe a regular system that admits one autonomous integral of motion generated by the corresponding Killing field of $G_{\alpha\beta}$ \footnote{Some extra cases admitting rheonomic integrals of motion are also explored, for more details see \cite{tsamp}}. However, one can notice that all three scalar field potentials belong to the same functional form, namely the form that makes the scaled mini--supermetric $\xbar{G}_{\alpha\beta}= U\, G_{\alpha\beta}$ flat for the specific value $F_{\mathrm{fixed}}(\phi)$ of the coupling function.

In this work, we use the reparametrization invariance which leads to the consideration of the scaled mini--supermetric $\xbar{G}_{\alpha\beta}$ as the crucial element describing the geometry of the configuration space and the dynamics of the system. Consequently, the demand for maximal symmetry does not fix the coupling function, but yields a relation between $F(\phi)$ and the scalar potential $V(\phi)$ \eqref{FVpar}. For the specific value $F_{\mathrm{fixed}}(\phi)$, treated in \cite{tsamp}, the potential $V(\phi)$ assumes the general functional form in which the three potentials given in that paper belong. Thus, what is considered as three different cases admitting one autonomous integral of motion in the study of the regular Lagrangian, is really one case admitting three autonomous integrals of motion in the actual (singular) cosmological system. Moreover, this is just a single case in the study of the singular Lagrangian \eqref{Lag}, since $F(\phi)$ is not fixed to an explicit functional form. Thus, the particular function $F_{\mathrm{fixed}}(\phi)$ is a choice, not a necessity for satisfying the demand of a flat the mini--superspace. The result is an infinite set of scalar tensor theories admitting the maximal number of autonomous, linear in the momenta integrals of motion. As we proved, even if one requires less symmetries, i.e. one or two autonomous charges, one is led to the case here examined: Each choice of $F(\phi)$, yields though \eqref{FVpar} the appropriate potential for a maximally symmetric (eventually flat) mini--superspace.

For all infinite cases that arise from the condition of maximal symmetry, we were able to acquire the general solution space for an arbitrary coupling function $F(\phi)$. This is not to be taken lightly; it means that the obtained sets \eqref{final_caseI} and \eqref{final_caseII} represent the general analytic solutions of \underline{every} scalar tensor, spatially flat FLRW cosmological theory that admits an autonomous, linear in the momenta symmetry. We also calculated all the physically relevant parameters and the effective energy--momentum tensor associated with the scalar field, which is seen to be mimicking a perfect fluid behaviour from Einstein's gravity's perspective.

We would like to emphasize that the correspondence between the scalar field and the perfect fluid we use is not the usual: The common practice is to identify the energy momentum tensor of the scalar field \eqref{Tij}  ($V(\phi)=0$) with the energy momentum tensor of a perfect fluid. Our line of thinking is to rewrite the field equations in the form $E_{ij}=T_{ij}^{\phi}$  and treat the rhs as an energy momentum tensor. This different approach is responsible for enabling us to arrive to physically meaningful results.

In order to exhibit the way the general relations can be used and to complete our analysis, we have given some specific examples: a) A solution for a particular choice of the coupling function, that satisfies all the energy conditions (apart from the strong energy condition) and whose behaviour considerably matches with many observational facts. b) For the shake of completeness, we investigated the case when the mini--supermetric is degenerate, the only instance that is not covered by the general theory. This led to a solution that is seen to be equivalent to the dust solution of Friedmann in the context of General Relativity. c) We also obtained the known solution of General Relativity for a perfect fluid with a constant equation of state parameter $w$. This happened by considering the case $F(\phi)=1$ (minimally coupled scalar field) and by a suitable choice of the parameters entering the effective energy--momentum tensor. Of course this is not a new solution, but it serves to exhibit that the general solution for an arbitrary $F(\phi)$ correctly correlates to Einstein's theory when one sets $F(\phi)$ to a constant.

Since the presented method is quite a general one, it would be interesting to apply it to a broader setting i.e. one could add an actual perfect fluid along with the scalar field or explore the possibility of the existence of two scalar fields or even in more general theory like Horndeski's \cite{Horndeski1974}.

\newpage

\appendix

\section{Conformal 2D metric with one Killing field}\label{2Dcase}
Let's assume that the 2D metric $h_{\alpha\beta}$ admits the Killing field $\xi=\partial_x$, then $h_{\alpha\beta}=h_{\alpha\beta}(y)$ and the line element assumes the form
\bal\label{2D_y}
\drm s^2=F_1(y)\drm x^2+2F_2(y)\drm x\, \drm y+F_3(y)\drm y^2.
\eal

If $F_1(y)=0$, then the metric \eqref{2D_y} has Lorentzian signature $\det(h_{\alpha\beta})=-F_2^2(y)$ and is flat, since it can be transformed as follows
\bal
\drm s^2&=2F_2(y)\drm y\left( \drm x+\frac{F_3(y)}{2F_2(y)} \,\drm y\right), & \non\\
&= \frac{4F_2^2(y)}{F_3(y)} \drm z \left(\drm x+\drm z  \right),   & \frac{F_3(y)}{2F_2(y)} \,\drm y=\drm z, \, y=r(z) \non\\
&=R(z)  \drm z \left(\drm x+\drm z  \right),   & \frac{4F_2^2(r(z))}{F_3(r(z))}= R(z)\non\\
&=R(z) \drm z \drm w, & w=x+z\non \\
&=\drm u \drm w, & R(z) \drm z=\drm u.
\eal

If $F_1(y) \neq 0$ and $F_2(y) = 0$, then the metric \eqref{2D_y} transforms (assuming $F_1(y)>0$)
\bal
\drm s^2&=F_1(y) \drm x^2+\epsilon \left(\sqrt{|F_3(y)|} \drm y\right)^2, &\non\\
&=F_1(y)\drm x^2+\epsilon\, F_1(y)dz^2, & \sqrt{\left|\frac{F_3(y)}{F_1(y)}\right|} \drm y=\drm z, y=r(z) \non\\
&=\Omega(z)\left(\drm x^2+\epsilon\, dz^2\right) & \Omega(z)=F_1(r(z))
\eal

Finally, if $F_1(y) \neq 0$ and $F_2(y) \neq 0$, then the metric \eqref{2D_y} transforms (assuming $F_1(y)>0$)
\bal
\drm s^2&=F_1(y)\left( \drm x^2+\frac{2F_2(y)}{F_1(y)}\,\drm x\, \drm y  \right)+F_3(y)\drm y^2, & \non\\
&=F_1(y) \left[  \left(  \drm x+\frac{F_2(y)}{F_1(y)}\, \drm y  \right)^2 - \frac{F_2^2(y)}{F_1^2(y)}\, \drm y^2 \right]+F_3(y)\drm y^2, & \non\\
&=F_1(y)\left[  \left(  \drm x+\drm z \right)^2 -\drm z^2  \right]+\frac{F_3(y) F_1^2(y)}{F_2^2(y)}\,\drm z^2, & \frac{F_2(y)}{F_1(y)}\, \drm y=\drm z, y=r(z) \non\\
&=F_1(y)\drm w^2+\left( \frac{F_3(y) F_1^2(y)}{F_2^2(y)}-F_1(y)  \right)\drm z^2, & w=x+z\non\\
&=S(z)\drm w^2+\underbrace{\left( \frac{F_3(y) F_1^2(y)}{F_2^2(y)}-F_1(y)  \right)}_{F_4(y)} \drm z^2, & S(z)=F_1(r(z)) \non \\
&=S(z)\drm w^2+R(z) \drm z^2, &  R(z)=F_4(r(z))\non \\
&=S(z)\drm w^2+\epsilon \left(\sqrt{|R(z)|}\drm z\right)^2, & \non\\
&=S(z) \drm w^2+\epsilon\,S(z)\drm u^2, & \sqrt{\left|\frac{R(z)}{S(z)}\right|}\drm z=\drm u, z=t(u)\non\\
&=\Omega(u)\left(  \drm w^2+\epsilon\, \drm u^2 \right), & \Omega(u)=S(t(u)).
\eal

\phantomsection
\addcontentsline{toc}{section}{References}


\providecommand{\href}[2]{#2}\begingroup\raggedright\endgroup

\end{document}